\documentclass[12pt,preprint]{aastex}

\shorttitle{Are High-Redshift Quasars Blurry?}
\shortauthors{Steinbring}

\input epsf

\def\plottwo#1#2{\centering \leavevmode
\epsfxsize=0.5\columnwidth \epsfbox{#1}
\epsfxsize=0.5\columnwidth \epsfbox{#2}}

\def\plotonetwothirds#1{\centering \leavevmode
\epsfxsize=0.666667\columnwidth \epsfbox{#1}}

\begin{document}

\title{Are High-Redshift Quasars Blurry?}

\author{Eric Steinbring\altaffilmark{1}}

\altaffiltext{1}{Herzberg Institute of Astrophysics, National Research Council Canada, Victoria, BC V9E 2E7, Canada}

\begin{abstract}
It has been suggested that the fuzzy nature of spacetime at the Planck scale may
cause lightwaves to lose phase coherence, and if severe enough 
this could blur images of distant point-like sources sufficiently that they do 
not form an Airy pattern at the focal plane of a telescope.
Blurring this dramatic has already been observationally ruled out by images 
from {\it Hubble Space Telescope} ({\it HST}), but I show that the underlying phenomenon
could still be stronger than previously considered. It is harder to detect, which may 
explain why it has gone unseen. A systematic search is made in archival {\it HST} images 
of among the highest known redshift quasars. Planck-scale induced blurring may be 
evident, but this could be confused with partially resolved sources.
\end{abstract}


\section{Introduction}\label{introduction}

Characterizing the microscopic properties of spacetime using images of distant sources was
first proposed in a Letter by \cite{Lieu2003}.
This compelling possibility is a consequence of a phenomenon 
likely to be found at the Planck scale, where 
lengths shrink to $l_{\rm P} \sim 10^{-35}$ m and time intervals diminish to $t_{\rm P} \sim 10^{-44}$ s.
Following \cite{Ng2003}, here distance measurements $l$
are uncertain by $\delta l \gtrsim l_{\rm P}(l/l_{\rm P})^{1 - \alpha}$ (with similar uncertainties for time), where 
the parameter $\alpha$ specifies different quantum gravity models.
A natural choice for $\alpha$ is 1, but ${1\over2}$ is also possible \citep{Amelino-Camelina2000}, 
and ${2\over3}$ is preferred \citep{Ng2002}, because it is consistent with the holographic 
principle of \citet{tHooft1993} and established black-hole theory \citep{Beckenstein1973, Hawking1975}.
An electromagnetic wave travelling a distance $L$ from source to observer would 
be continually subjected to these random spacetime fluctuations, which following \cite{Ng2003} 
leads to a cumulative statistical
phase dispersion of
$$\Delta \phi_0 = 2 \pi a_0 {l_{\rm P}^{\alpha}\over{\lambda}} L^{1-\alpha}, \eqno(1)$$
where $\lambda$ is the observed wavelength, and $a_0$ is close to unity.
This presents a means of
detection in an image of a point source.
\cite{Ragazzoni2003} point out that $\Delta \phi_0$ can
be interpreted as an apparent angular broadening of the source, and over a sufficient distance might
grow enough to obliterate the diffraction pattern at the focal plane 
of a telescope. 

Because this has not been seen in {\it Hubble Space Telescope} ({\it HST}) data, Lieu et al. and other 
authors \citep{Ragazzoni2003, Ng2003} have 
concluded that the effect, if present, is too weak to be observed. 
In Section~\ref{limits} I expand on previous work to account for the shorter
wavelength of emitted light, which predicts stronger blurring. But observing this is still 
problematic, because it must be disentangled from a partially resolved
source. A test is outlined in Section~\ref{selection}
which makes use of the best available archival data, a {\it HST} snapshot survey of $z>4$ quasars.
These data are discussed in Section~\ref{reductions}, after which a 
new analysis in the context of blurring by Planck-scale effects is presented in Section~\ref{results}.

\section{Limits on Blurring Induced by the Planck Scale}\label{limits}

\begin{figure}
\plotonetwothirds{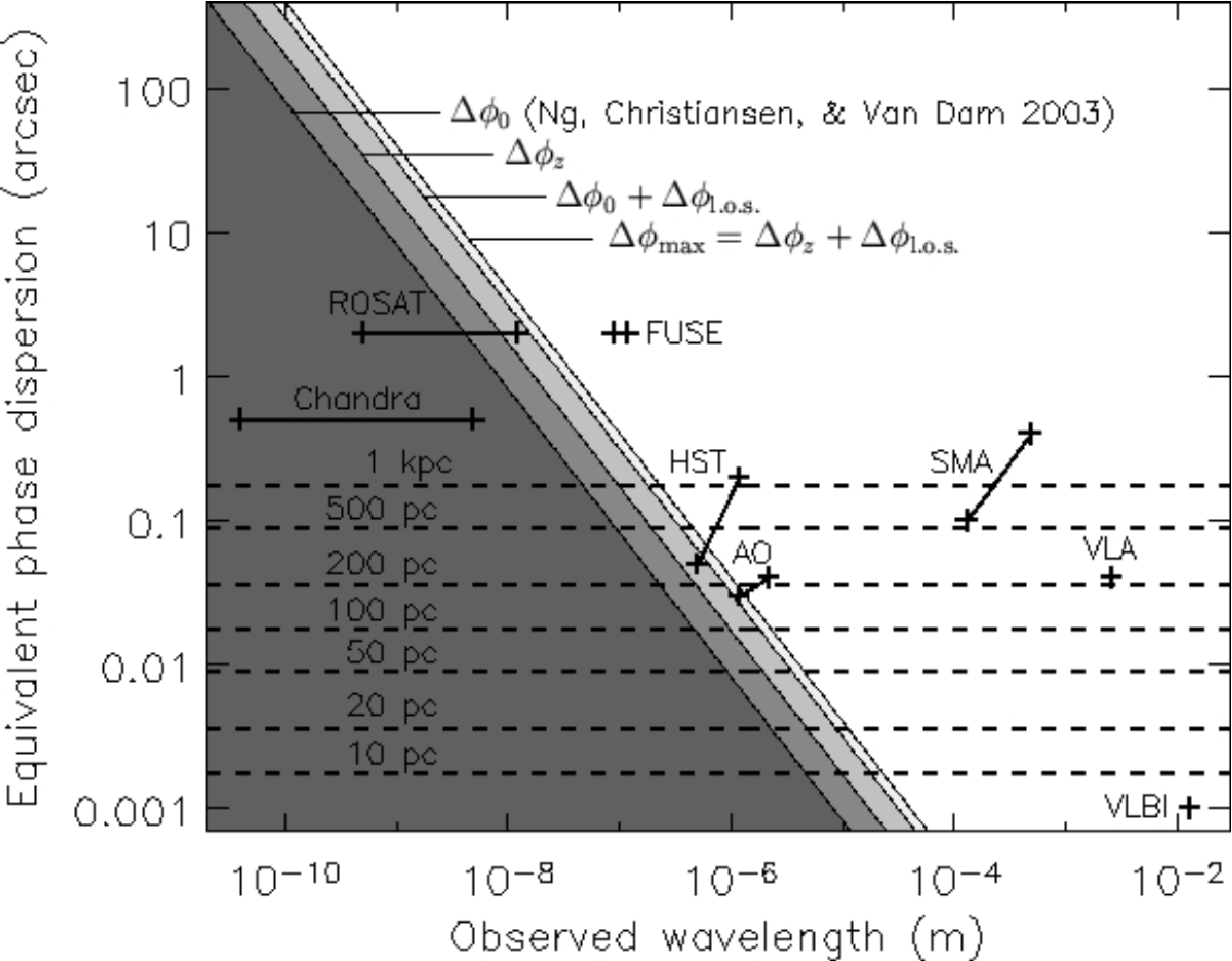}
\caption{Accumulated phase dispersions in arcseconds as a function
of wavelength for a point source with $z=4$, according to equations 1 and 3 with 
$\alpha={2\over3}$ and $a_0=1$. By comparison, the resolution limts of some
telescopes are indicated: Chandra X-ray Observatory (Chandra), Roentgen Satellite (ROSAT), 
Far Ultraviolet Spectroscopic Explorer (FUSE), adaptive optics on an 8-m-class telescope (AO), Smithsonian Submillimeter Array (SMA), 
Very Large Array (VLA), and Very Long Baseline Interferometry (VLBI). The sizes of extended objects of 
various physical diameters are shown 
as dashed lines. Shaded regions bordered by equations 1 and 3 (and two intermediate subregions formed by either excluding or combining their terms) are potentially precluded from observation.}
\label{figure_multiwavelength}
\end{figure}

Equation 1 assumes that the emission and detection wavelengths are the same, and so 
represents a minimal estimate for blurring.
For cosmological distances, photons left the source at the shorter wavelength $\lambda\over{1 + z}$, for which the 
blurring action of the Planck scale should be stronger.
Consider the following:
$$\Delta \phi_{\rm max} = 2 \pi a_0 (1 + z) {l_{\rm P}^{\alpha}\over{\lambda}} L^{1 - \alpha} = (1 + z) \Delta \phi_0. \eqno(2)$$
One might call this a maximal estimate, because it assumes the bluest 
possible photons propagate a distance $L$ and are then detected at wavelength $\lambda$.
It is instructive to write equation 2 in a different form. By taking its differential with 
respect to $z$, it is possible to rewrite it as 
$$\Delta \phi_{\rm max} = 2 \pi a_0 {l_{\rm P}^{\alpha}\over{\lambda}} \Big{\{}\int_0^z L^{1 - \alpha} dz + {{(1-\alpha)c}\over{H_0 q_0}} \int_0^z (1+z) L^{-\alpha} \Big{[}1 - {{1 - q_0}\over{\sqrt{1 + 2 q_0 z}}}\Big{]} dz \Big{\}}$$
$$ = \Delta \phi_{\rm l.o.s.} + \Delta \phi_z. ~~~~~~~~~~~~~~~~~~~~~~~~~~~~~~~~~~~~~~~~~~~~~~~~~~~~~~~~~~~~\eqno(3)$$
where the luminosity distance $L$ is given by
$$L = {c\over{H_0 q_0^2}}\Big{[}q_0 z - (1 - q_0)(\sqrt{1 + 2 q_0 z} - 1)\Big{]} \eqno(4)$$
and $q_0={{\Omega_0}\over{2}} - {{\Lambda c^2}\over{3 H_0^2}}$ is the deceleration parameter.
By comparison with equation 1, the first integral can be recognized as the observed phase dispersion of 
photons of wavelength $\lambda$ arriving from all redshifts up to the source.
In practice these could conceivably come from intervening sources along the line of sight, hence the label $\Delta \phi_{\rm l.o.s.}$.
 The second integral must then be the remaining phase dispersion $\Delta \phi_z$ associated exclusively with 
photons redshifted to the observer's wavelength. The latter scenario seems the more plausible.
This is plotted in Figure~\ref{figure_multiwavelength} for a $z=4$ point source as a function 
of observed wavelength using the standard LCDM cosmology ($\Omega_\Lambda=0.7$, $\Omega_{\rm M}=0.3$, and $H_0=70~{\rm km}~{\rm s}^{-1}~{\rm Mpc}^{-1}$)
assumed throughout this paper. Three other possibilities are also plotted: minimal blurring $\Delta \phi_0$, the combination 
of minimal blurring of the source plus that of intermediate sources along its line of sight $\Delta \phi_0 + \Delta \phi_{\rm l.o.s.}$, and maximal blurring $\Delta \phi_{\rm max}$, for a total of four.
Following \cite{Ng2003} I use $\alpha={2\over3}$ and
$a_0=1$. Reducing (increasing) $a_0$ shifts these lines in parrallel 
towards lower left (upper right) in this plot, as does increasing (decreasing) $\alpha$. 
The most dramatic changes come with variation in $\alpha$, by virtue of it appearing as an exponent in equations 1 and 3.

Can blurring be constrained by current observations?
Figure~\ref{figure_multiwavelength} illustrates at what wavelengths an answer may be most convincing.
The highest spatial resolutions of some current telescopes
are indicated.
Chandra routinely observes sources to a resolution limit better than that imposed
by either equations 1 or 3 for $\alpha={2\over3}$. But this conflict is easily avoided, by either 
suitably increasing $\alpha$ (0.7 would be sufficient) or reducing $a_0$ - or both.
 This flexibility is allowed because current telescopes operating shortward of the 
optical (including Chandra) do not form diffraction-limited images, and so it may be more difficult to 
distinguish blurring induced by the
Planck scale from simply an unresolved source such as a galaxy. The dashed lines indicate 
the angular size $\Delta \theta$ of 
extended $z=4$ sources for a range of 
physical diameters. 
Thus, it may be best to look for blurring in the optical.
{\it HST} probes a regime predicted to be blurry, which is indicated
by the shaded regions. It can also resolve sources confirmed to be physically small at
other wavelengths. For example, if $\alpha={2\over3}$ and subject to the maximum blurring limit, no 
$z=4$ source should appear sharper 
than about 0\farcs1 across at 800 nm, broad enough to affect the {\it HST} diffraction pattern.
An object with $z=4$ would need to be 500 pc across to mimic this level of broadening.
This is smaller than a typical galaxy, but not quasars. Even if it were a factor of about 2.5 less,
{\it HST} could still marginally discriminate between Planck-scale-induced blurring and the effects of a 
200 pc source. But a $\sim 25$-m space telescope would be needed to rule out the minimum set by equation 1, by 
resolving objects as small as 50 pc.

\section{Observational Plan and Sample Selection}\label{selection} 

\begin{figure}
\plotonetwothirds{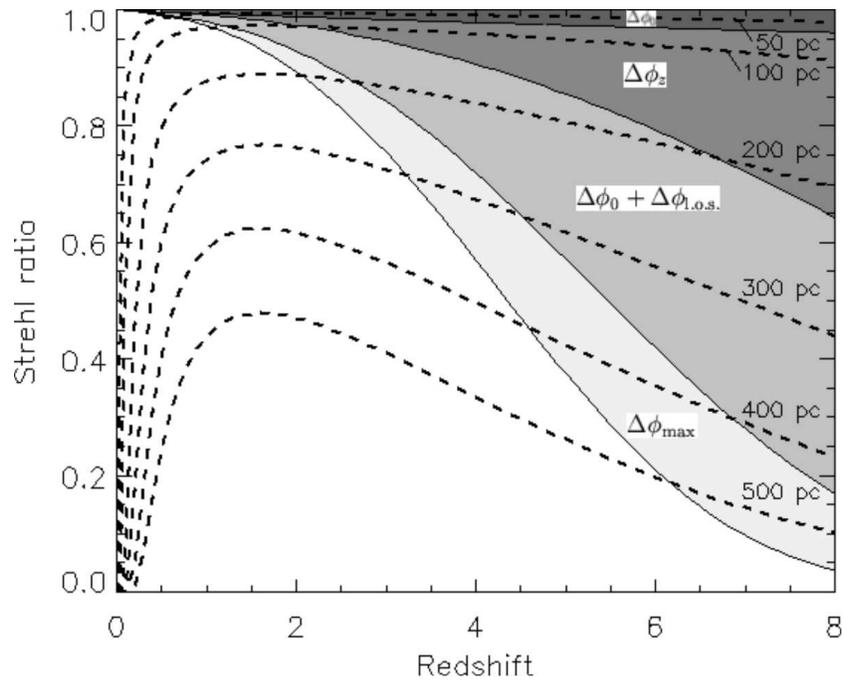}
\caption{Precluded Strehl ratio for blurred point sources (shaded regions) and 
predicted Strehl ratio for partially-resolved extended sources (dashed lines) as a 
function of redshift for {\it HST} observing at 800 nm. Shaded regions have the same
meanings as in Figure~\ref{figure_multiwavelength}.}
\label{figure_models}
\end{figure}

\cite{Ragazzoni2003} have already confirmed that the most severe blurring discussed in Section~\ref{limits} is 
not seen in {\it HST} images of a $z=5.4$ Hubble Deep Field galaxy. It is worth considering how weak
blurring could be and still be ruled out for {\it HST}.
Ragazonni et al. point out that small perturbations of the phase should only cause a drop in image Strehl ratio $S$.
This is the ratio of the image peak to that of the diffraction spike of an unabberated telescope.
In this case, and where $\Delta \phi$ is comparable to the telescope optical aberrations, the Marechal
approximation to the Strehl ratio applies, and 
$$S\approx\exp{\Big{[}-\Big{(}\Delta \phi {D\over{\lambda}}\Big{)}^2\Big{]}}. \eqno(5)$$
Figure~\ref{figure_models} is a plot of point-source Strehl ratio as a function
of redshift for an unaberrated telescope with $D=2.4$ m observing
at $\lambda=800$ nm. The shaded regions have the same meaning as in Figure~\ref{figure_multiwavelength}, again
plotted for $\alpha={2\over3}$ and $a_0=1$.
If the object is not a true point source but is actually extended, equations 1 and 3 do not strictly apply. 
In this case the Strehl ratio of the image would also drop due to the partial
resolution of the source. For an object of angular size $\Delta \theta$ close to the 
diffraction limit of the telescope one would expect a Strehl ratio of
$$S\approx \exp{\Big{[}-\Big{(}\Delta \theta {D \over{\lambda}}\Big{)}^2\Big{]}} \eqno(6)$$
in the absense of Planck-scale effects. The dashed lines in Figure~\ref{figure_models} represent equation 6 for 
various source diameters.

Judging from Figure~\ref{figure_models}, to test if high-redshift sources are indeed blurred, one could look at a 
large 
homogeneous sample of compact high-redshift objects imaged with {\it HST}.
In principle, the test is simple: Is there 
a downward trend in Strehl ratio with increasing source redshift? 
In practice, two factors must be well controlled. First, the sample must be of
uniformly small sources of very high redshift.
Figure~\ref{figure_models} indicates that for any Planck-scale effect to 
be detected with {\it HST} the sample must extend to redshifts well beyond $z=4$.
Second, the point-spread function (PSF) must be
well characterized because the Planck-scale-blurring signal is 
relative to the expectation for diffraction-limited imaging.

As it turns out, there is already at least one archival 
{\it HST} dataset that meets these strict critera: the Advanced Camera for Surveys (ACS) High-Resolution Channel 
(HRC) snapshot survey of SDSS quasars.
This was a program of imaging to detect lensed companions among a sample of $4<z<6.3$ quasars.
None were found despite the excellent spatial resolution afforded by ACS HRC. 
Because depth was not as important as spatial resolution, these images are relatively shallow, which
should minimize confusion by any extended host galaxy. Also, \cite{Kaspi2005} suggest that none of the
SDSS quasars should have a broad-line region larger than a few parsecs, which would make equation 6 negligibly 
less than unity for {\it HST}.
Thus, this sample should satisfy the first condition. That they also satisfy the second is
a property of both {\it HST} PSF stability and the Nyquist-sampled ACS HRC pixel scale. Neither the
Space Telescope Imaging Spectrograph (STIS) nor the Wide Field and Planetary Camera 2 (WFPC2) can
fulfill this second constraint.

\section{Data and Reductions}\label{reductions}

The archival {\it HST} ACS snapshot survey
of Sloan Digital Sky Survey (SDSS) sources (Proposal: 9472, PI: Strauss) includes all 
SDSS quasars with $z>4$; some of which have $z>6$. They were obtained in snapshot mode 
through either the F775W filter (central wavelength 761 nm, 95 sources, $3.9<z<5.4$) or
F850LP (869 nm, 4 sources, $5.8<z<6.3$). All were imaged with the HRC, which has a 
pixel scale of 0\farcs0246.

These data have already been discussed in detail in \cite{Richards2004, Richards2006}, and
those reductions are closely followed here.
Pipeline-processed images were downloaded from the {\it HST} archive. This provides
standard corrections for overscan, bias, and dark-current, flatfield division, bad-pixel masking, 
cosmic-ray removal, and photometric calibration.
The position of the quasar was determined to the nearest pixel using the IRAF\footnote[1]{IRAF is 
distributed 
by the National Optical Astronomy Observatory, which is operated by the Association of Universities
for Research in Astronomy, Inc., under cooperative agreement with the National Science Foundation.} 
program IMEXAM.  Next, the Tiny Tim (v6.3) program \citep{Krist1995} was used to 
generate an appropriate PSF for each. Due to the strong
positional dependence of the HRC field distortion this produces a more accurate
PSF than can be determined by images of stars. Color dependence 
is also properly accounted for, by inputing
a spectral energy distribution (SED) and convolving this with the {\it HST} filter curve.
Synthetic PSFs were generated for each quasar using a redshifted template SED based on
a composite SDSS QSO spectrum \citep{VandenBerk2001}. By shifting in the Fourier domain, 
the sub-pixel position and peak of the PSF were allowed to float 
relative to the image, and minimized based on the sum of the square of residuals. The result is 
an unique PSF for each quasar, correct for its detector position, filter bandpass, and SED.
 
All of the images were then co-aligned to the center of the nearest pixel.
Figure~\ref{figure_images} shows the result for SDSS J0836+0054 ($z=5.82$), which will serve as an example.
Each box is 3\arcsec $\times$ 3\arcsec~with the same orientation as in \cite{Richards2004} (their Figure 2).
All images have the same grey-scale stretch.
The non-circular first Airy ring is clearly evident, and further rings would be visible in 
a harder stretch. The diffraction pattern of a circularly symmetric $2.4$-m diameter pupil 
with a $0.7$-m central obstruction is also shown. Note that this excludes any aberrations 
associated with the ACS camera, which explains its circular symmetry. The same SED as the quasar was used. 
It is relative to this ideal diffraction pattern that - by definition - the Strehl ratio of the quasar and PSF is 
to be determined. 
Next to this is the PSF residual (the model residual will be discussed later, in Section~\ref{results}). This residual 
is robust against variation in the template SED, both in relative
line strengths and choice of continuum power law.
It is also, as can be expected, comparable to that 
obtained by Richards et al. and reveals no obvious host galaxy or lensed component. But underlying effects
due to microlensing cannot be ruled out, although it is not clear what effect they may have.
Results are similar for the other sources.

Strehl ratio was then measured, which is not reported by
\cite{Richards2004, Richards2006}.
Each image and PSF was normalized to a flux of unity based on synthetic aperture photometry, and the height of its 
peak determined 
relative to the ideal diffraction pattern. The accuracy of the final
measurement is to within a few percent, which is the level of photometric uncertainty.

\section{Analysis and Results}\label{results}

Figure~\ref{figure_strehl} is a plot of Strehl ratio; crosses indicate the PSF. 
The unfortunate division between the F775W and F850LP data at $z=5.5$ is evident.
Even if one ignores this, a clear downward trend with increasing source redshift can be seen, despite some scatter in the data.
Overplotted is a linear least-squares fit to these, which has slope of $-0.05\pm0.01 z$ 
(1-$\sigma$ errors). 
That some of the lowest redshift quasars have Strehl ratios higher than
the PSF indicates that the limits of the telescope have been reached here. The
differences of a few percent are consistent with the uncertainties in the measurements.
Thus, the decrement below the PSF Strehl ratio for the $z>5$ redshift sources is probably real.
The situation for the $z>5.5$ sources is more secure. Limits predicted by equations 1 and 3 assuming $\alpha={2\over3}$ and $a_0=1$
are overplotted.
The break is due to the change in filter effective wavelength, from 761 nm to 869 nm.
For this choice of $a_0$ and $\alpha$, both $\Delta \phi_{\rm max}$ and $\Delta \phi_0 + \Delta \phi_{\rm l.o.s.}$ are clearly ruled out.
But blurring associated with just redshifted sources is not. It is striking how closely the maximum Strehl ratio
follows this curve.
And of those quasars for which the PSF should be sharp enough to allow it, none are found in the region that 
it borders.
If this is correct, it places a tight constraint on $\alpha$.
Even if $a_0=10$, $\alpha$ need only grow
by 2\% (to 0.68) and still be in agreement with the data.
Although also possible, this result is not easily explained by
the size of the sources alone. It would seem that the intrisic sizes of these quasars (either their narrow-line or broad-line regions - or both) would need to occupy a very narrow range for this to happen, between about 200 pc and 300 pc.

\begin{figure}
\plotonetwothirds{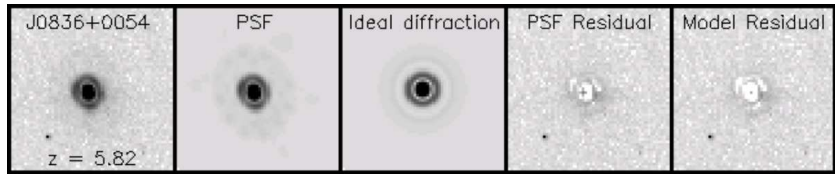}
\caption{Images of the quasar, PSF, ideal diffraction pattern, and residuals for one of the QSOs in the sample.
Results are similar for the others.}
\label{figure_images}
\end{figure}

\begin{figure}
\plotonetwothirds{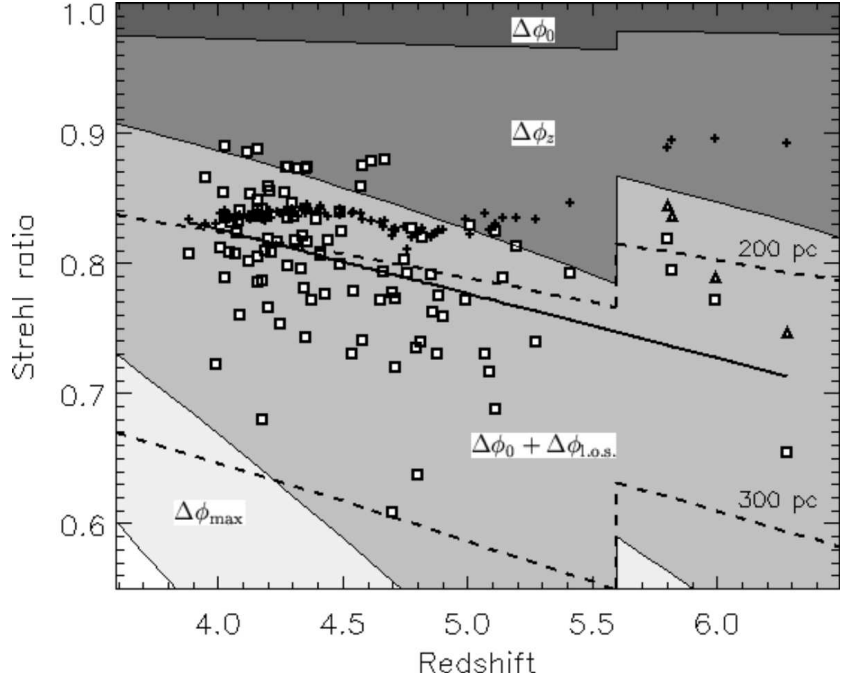}
\caption{A plot of Strehl ratio (open squares) for the sample. The crosses are the PSF for each 
quasar, giving an optically corrected Strehl ratio via the Tiny Tim software.
There is a noticable trend towards lower observed Strehl ratio with increasing redshift. A linear
least-squares fit to all data is overplotted. No correction for the change in filter at $z=5.5$ has been
made. Models of Planck-scale induced blurring are shown; the shaded regions have the same meanings as 
in Figures~\ref{figure_multiwavelength} and ~\ref{figure_models}.
The break is due to the shift in filter central wavelength. Open triangles indicate the ``blurred" quasar model
of Section~\ref{results}. The region bordered by $\Delta \phi_z$ is avoided, and so cannot
be ruled out by the observations.}
\label{figure_strehl}
\end{figure}

\begin{figure}
\plottwo{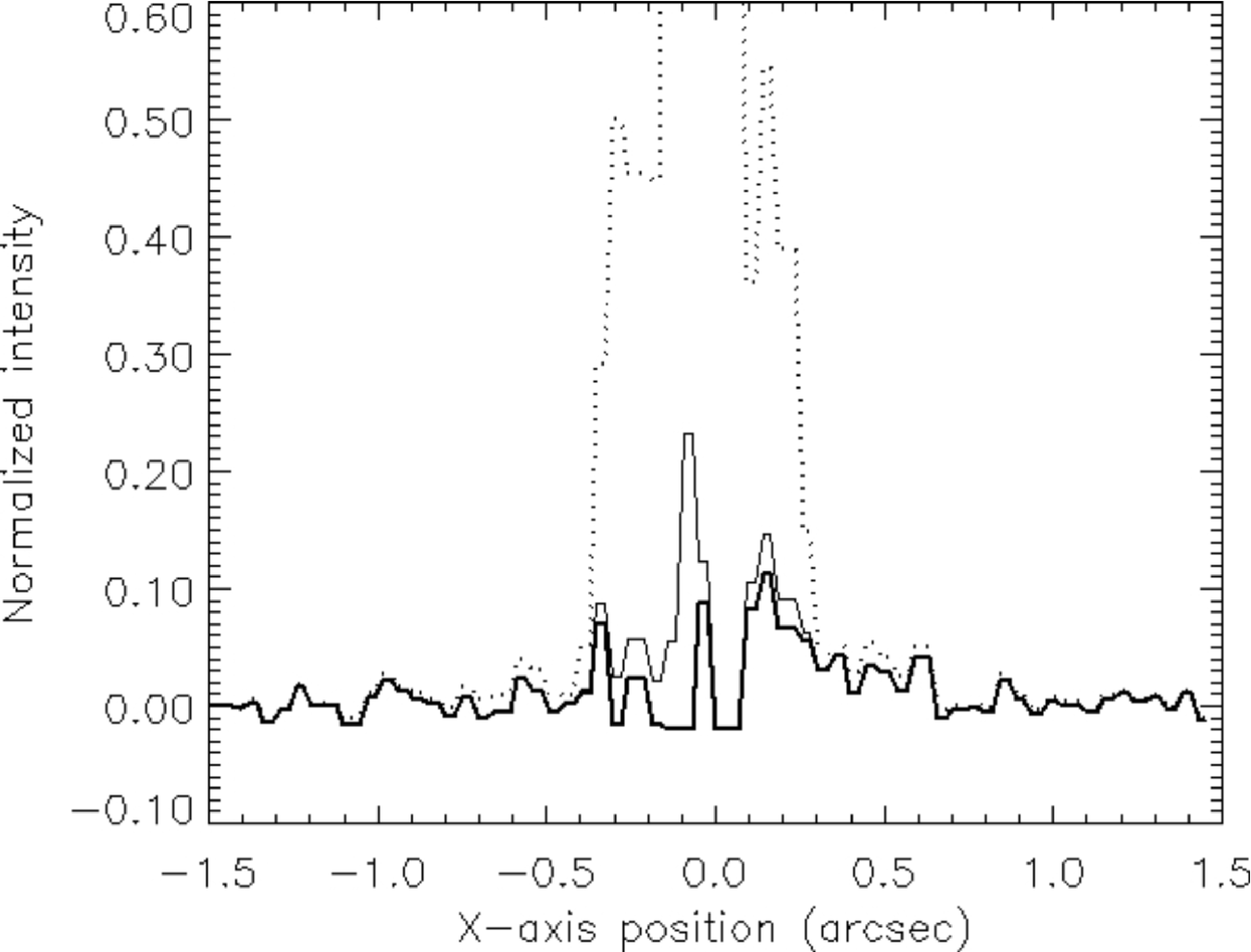}{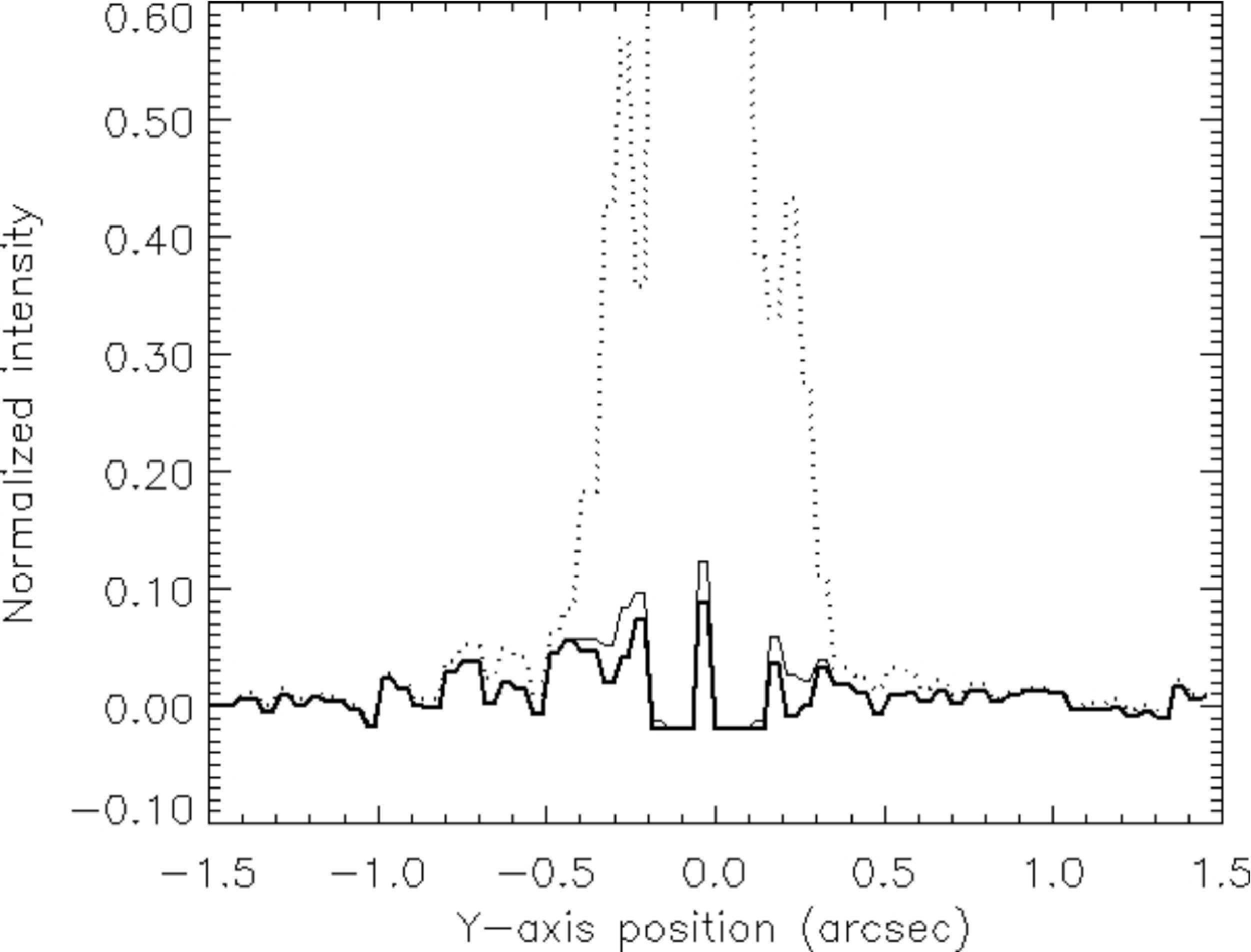}
\caption{Slices along the x and y axes of the same quasar shown in Figure~\ref{figure_images} (dotted line). Overplotted are
the PSF residual (thin solid line) and model residual (thick solid line).  All are normalized
to the peak intensity in the quasar image. The scale has been restricted to 0.6 and below to better display the 
residuals. The model, although
still not perfect, is a better fit to the quasar.}
\label{figure_profiles}
\end{figure}

In a further simple test of the internal consistency of the $z>5.5$ results, a simple ``blurred" quasar
model was generated.  Guided by the good fit of $\Delta \phi_z$ in Figure~\ref{figure_strehl}, the existing PSF was 
cropped slightly below
its peak, at a cutoff given by ${\rm cutoff}={\rm peak}\times\exp{\Big{[}-\Big{(}\Delta \phi_z {D\over{\lambda}}\Big{)}^2\Big{]}}$. A
value $\lambda = 869$ nm was used, for $D=2.4$ m.  
This naive model is indeed a better fit to the observed Strehl
ratio, indicated by the open triangles in Figure~\ref{figure_strehl}. The effects on the residual are benign, which 
can be seen in Figure~\ref{figure_images}. This is also evident in slices along the x and y axes, which are plotted 
in Figure~\ref{figure_profiles}. This confirms that the blurred quasar model is a better fit than the original PSF, 
accomplished without adversely affecting the residual.  This is reassuring, as it demonstrates that these results are not 
in conflict with those of \cite{Richards2004}.

In summary, 
I have searched for blurring induced by the effects of the Planck scale in {\it HST} images of 
high-$z$ quasars. Although blurring may be seen, if real, it is just at the observable threshold.
The test here is more sensitive than previous ones, which only looked for compact sources lacking diffraction rings. 
The shorter wavelength of emitted light has also been accounted for, which if correct, 
restricts $\alpha$ to be 0.68 even if $a_0$ is as big as 10.

The implications of blurring are significant, so it is worth looking for.
A true detection could point to a successful quantum gravity theory.
But detection is elusive because 
the signal is confused with that of a partially resolved source.
It is hoped that this work will encourage further tests for the 
effect with {\it HST}. A first step would be to 
re-observe the four $z>5.5$ quasars in this sample with the HRC and the F775W filter.
Observations of higher-$z$ quasars (as they become known) could ultimately confirm or refute
the current result.

\clearpage

\clearpage

\clearpage

\clearpage

\end{document}